\documentclass[epj]{svjour}
\usepackage{graphics}
\begin{document}
\hyphenation{mono-mers}
\hyphenation{bran-ches}
\title{Monte-Carlo simulations of star-branched polyelectrolyte micelles}
\author{M. Roger\inst{1}\thanks{\emph{email: roger@drecam.saclay.cea.fr}},
P. Guenoun\inst{1}, F. Muller\inst{1}, 
L. Belloni\inst{2} and M. Delsanti\inst{2}.}

\institute{DRECAM, Service de Physique de l'Etat Condens\'e,
 CEA Saclay, 91191 Gif sur Yvette Cedex, France. \and 
DRECAM, Service de Chimie Mol\'eculaire, CEA Saclay, 91191 Gif sur Yvette
Cedex, France.}
\date{Received: date / Revised version: date}
\abstract{The concentration profiles of monomers and 
counterions in  star-branched polyelectrolyte micelles
are calculated through Monte-Carlo simulations, using the simplest
freely-jointed chain model.
We have investigated the onset of different regimes
corresponding to the spherical and Manning condensation
of counterions as a function of the strength of the Coulomb coupling. 
The Monte-Carlo results are in  fair
agreement with the predictions of Self-Consistent-Field analytical models.
 We have simulated a real system of diblock copolymer micelles
of (sodium-polystyrene-sulfonate)(NaPSS)-- (polyethylene-- propylene)(PEP)
with f=54 hydrophilic branches of $N=251$  monomers at room temperature in 
salt-free solution and compared the
calculated form factor with our neutron-scattering data.}

 % end of PACS codes
\authorrunning{M. Roger et al.}
\titlerunning{Monte-Carlo simulations of star-branched polyelectrolytes}
\maketitle
\section{Introduction}
\label{intro}
Charged diblock copolymers with a long and fully
charged hydrophilic part and a short hydrophobic head are attracting a great
attention.
In water, their hydrophobic cores tend to aggregate and they form colloidal 
solutions of star-branched spherical micelles. 
These physical systems have  many  applications in surface adhesion, 
stabilization of colloidal solutions and are also used in drug delivery
processes. 
Due to the presence of long-range Coulomb interactions,
simple scaling arguments are generally unsufficient to describe
their behavior. Quantitative predictions concerning
the concentration profiles of monomers and counterions have been
obtained through Self-Consistent-Field (SCF) 
models  \cite{Borisov1,Borisov2,Borisov3}.
Numerical simulations can also provide useful information. The
Molecular Dynamics (MD) approach has been widely used in the study of neutral
star-branched micelles \cite{Kremer1} and charged linear 
polyelectrolytes \cite{Kremer2,Kremer3}. Recently Jusufi, Likos and
L\"owen (JLL) \cite{Likos1,Likos2} applied it to star-branched polyelectrolyte
micelles. They studied, at relatively low linear charge densities of the
chains, the monomer and counterion profiles
of one micelle and  the effective interaction between two micelles. 
They also proposed analytical approximations to be compared with the results 
of MD simulations.

We focus here our attention on the problem of one isolated micelle, 
in a  different strongly-charged regime, 
which is relevant for some dilute solutions
of (sodium-polystyrene-sulfonate)(NaPSS)-- (polyethylene-- propylene) (PEP)
that have been investigated through neutron 
scattering in our laboratory \cite{Muller1}.

\noindent
We use the Monte-Carlo  method which, through a stochastic
approach, should provide the same information as the Molecular Dynamics. 
Our final goal is to reach  real-system sizes  like stars
with $f=54$ arms of $N=251$ monomers (27108 particles), corresponding  to
experimental conditions and compare with our neutron-scattering data,
without any adjustable parameter. 
As a prelude, we first  present a systematic investigation
of smaller micelles within a more general theoretical framework and compare
our numerical Monte-Carlo results with the predictions of scaling
theories and analytical approximations.
% \cite{Borisov1,Borisov2,Borisov3,Likos1,Likos2}.

After the description of our model
and Monte-Carlo method (section 2), we study in section 3 the evolution 
of the structure of a micelle with $f=27$ arms of $N=130$ monomers 
 as a function of the following dimensionless parameter 
measuring the strength of the Coulomb interaction:
\begin{equation}
\zeta =|z_i|\xi=|z_i||z_m|l_B/a
\end{equation}
 where
\begin{equation}
l_B=(e^2/4\pi\epsilon\epsilon_0)(1/k_BT)
\end{equation}
 is the Bjerrum length, $\xi=|z_m|l_B/a$ represents 
the ``Manning parameter'' \cite{Manning} which characterizes the linear 
charge density of polymers; $a$ is the distance between consecutive charged 
monomers, $T$ is
the temperature, $\epsilon$ the dielectric constant of the solvent and 
$z_m$, $z_i$ represent the valence of monomers and counterions respectively.
Throughout our study $z=-z_m=z_i>0$ and all monomers are charged.

The Bjerrum length is $l_B\approx 0.71~nm$ in water at room temperature and
it can be varied through the temperature and dielectric
constant. The monomer length is usually of the order of a fraction of a
nanometer. $\zeta$ is approximately 3 for sodium polystyrene sulfonate 
in water. It can be increased by choosing other solvents with lower
dielectric constants and$/$or by taking polyvalent monomers 
and counterions ($z>1$).
Higher values of $\zeta$ are thus easily accessible. Although,
in our fully charged model  small
values of $\zeta$, much less than 1, are at present experimentally 
unrealistic, it is nevertheless interesting to extend our theoretical
investigations down to that range.

 When the thermal energy
is much larger than the Cou-lomb interaction $(\zeta<<1)$, most counterions
are outside the micelle corona. This corresponds to the 
``unscreened regime'' \cite{Borisov1}.
Above some threshold
value $\zeta_{sph}\approx 1/f$, a spherical condensation of counterions
inside and in the vicinity of the corona is expected
 \cite{Borisov1,Borisov2,Borisov3,Luc1}. 
In this so-called ``osmotic regime'' \cite{Borisov1,Borisov2,Borisov3},
the spherically condensed counterions lower the interaction between branches.
 Both unscreened and osmotic regime have
been extensively studied through mean-field
analytical models \cite{Borisov1,Borisov2,Borisov3,Pincus1}.
 Finally, when the ``Manning criterion'' \cite{Manning} 
$\xi_M\approx 1/z_i$, i.e $\zeta_M\approx 1$, 
is reached, a ``cylindrical condensation''
of part of the counterions around the branches is superimposed to the 
spherical condensation. 
Scaling and
SCF theories \cite{Borisov1,Borisov2,Borisov3,PGDG} have predicted a
linear extension of the branches with corona radius $R_c$ proportional
to $N$ and monomer density varying roughly as $1/r^2$, where $r$ represents
the radial distance to the center of the micelle. 
This has justified a ``rod-like'' picture of the branches. This picture
has been  confirmed by clear experimental evidence of the stretching of the
chains \cite{Guenoun}. For highly charged polymers,
a Poisson-Boltzmann analysis \cite{PB}
has been proposed to interpret the distribution of counterions around the
rod-like branches in terms of a ionic condensation \cite{Manning,Luc2}. 

In the three latter regimes, we have compared the results 
of our simulations to the expectations from approximate analytical theories.
The simulation of a real system of (sodium-polystyrene-sulfonate) (NaPSS)-- 
(polyethylene-- propylene)(PEP) with $f=54$ branches of $N=251$ monomers,
is described in section 4, with a direct comparison to our neutron-scattering
data \cite{Muller1}.

\section{The Model and Monte-Carlo method}

\subsection{The model}
We use the simple ``freely-jointed chain'' (or ``pearl necklace'') model.
Each monomer is represented by a hard core of diameter $\sigma_m=a$. 
The angle between two successive monomer-monomer segments
 is free. We consider that the hydrophilic chain is
fully ionized in water and  each monomer carries a charge $(z_me)$. This is
relevant for sodium- polystyrene- sulfonate, in our experimental conditions,
where the degree of ionization is generally of order 0.95.
The counterions are modeled by hard spheres of diameter $\sigma_i=a/2$,
 they carry a charge $(z_ie)$. For sodium- polystyrene- sulfonate
$z_i=-z_m=1$.

One isolated micelle and the counterions are treated in the
 ``primitive cell model'' i.e. they are confined inside 
a hard-wall cube of edge $A$,
surrounding the center of the micelle. The dielectric constant is assumed to
be the same inside and outside the cubic cell.
This approach is relevant for low concentrations with respect to the
overlap concentration $c^*$ of micelles. 
The weak dependence of our physical results
on the size $A$ of this cubic box (provided it is larger than the
diameter of the micelle) means that  a more sophisticated
treatment of Coulomb interactions (e.g. periodic boundary conditions and
Ewald sums) is not required here.  

The hydrophobic core is modeled by a hard sphere of
 radius $R_{core}$ of order
5 to $20a$. The hydrophilic parts are attached at fixed points chosen
on this sphere. 
Two distributions of these fixed points have been considered:
\begin{itemize}
\item
{ a random distribution}
\item
{ a regularized distribution with approximate equidistance between 
first-neighbor pairs.} 
\end{itemize}
\noindent
The results obtained with these different distributions are practically
identical and it is thus not useful to perform mean values of
Monte-Carlo results over different realizations of
random distributions of points on the sphere.

\vskip .2truecm

{ All distances are expressed in units of the monomer length $a$ and
 the strength of the Coulomb interaction compared to $k_BT$ is measured by
the dimensionless parameter $\zeta$ defined in equation (1).}
 
\subsection{Monte Carlo moves}

For a counterion, an elementary move is chosen at random in a cube of edge
 $\delta x\approx 10a$ centered around it.

For polymers we use the ``pivot algorithm'', 
which has been shown to satisfy ergodicity \cite{pivot}.
A monomer $P$ is chosen at random, and a unit vector $\hat u$ is 
taken randomly 
in some solid angle $\delta\Omega$ centered about 
the axis defined by $P$ and the next monomer
$(P+1)$. A rigid rotation of random angle around $\hat u$ is applied 
to the part of the polymer going from $(P+1)$ to its end.
However, beyond Manning condensation $(\zeta >1)$, such a pivot algorithm
involving only the monomers becomes inefficient since many counterions
are localized around each polymer and the cost in 
energy to move a group
of monomers away from these counterions prohibits any Monte-Carlo move.

To circumvent this problem, we have used the ``partially clothed'' pivot 
algorithm proposed by Gordon and Valleau \cite{Valleau}.
 Having chosen a pivot $P$ and
the corresponding end part of the polymer, we look for all counterions at a
distance less than $d$ of this polymer part (the distance from the polymer
is defined as the distance to the closest monomer). Each of these
counterions is rotated in block with the part of the polymer, with a
probability $p\le 1$. The detailed-balance equation is satisfied if, in the
Metropolis scheme, we take as the
probability of acceptance of the Monte-Carlo move \cite{Valleau}
\begin{equation}
P_{acc}=min\{1,(1-p)^{n_{new}-n_{old}}\exp{[-(U_{new}-U_{old})/k_BT]}\},
\end{equation}
where $n_{new}$ and $n_{old}$ are the number of counterions at a distance
less than $d$ after and before the temptative rotation. $U_{new}$ and
$U_{old}$ are the Coulomb energies after and before the trial move.

In practice,  we have
observed  that ``clothed pivots'' have a reasonable probability to
be accepted in the Metro-polis scheme  if $p>0.9$. Since
we did not observe  very important differences in the convergence to
equilibrium for $0.9<p\le 1$,  we chose 
the value $p=1$ which makes the algorithm simpler and the 
calculation shorter. 
(The choice $p=1$ has also been made by Lobaskin and Linse \cite{Lobaskin} in
similar ``cluster moves'' of macro-ions surrounded by a shell of small
counterions).
The maximum ion-polymer distance for clothed pivots 
has been optimized  to $d=4a$.

To facilitate stretching and contraction of the polymer, we have also
applied some small ``clothed block translations'' of parts of polymer in the
following way. A monomer $P$ is chosen at random. The next monomer
$(P+1)$ is moved from $\vec r_{(P+1)}$ to $\vec r'_{P+1}$ within a 
solid angle $\delta\Omega$
centered around $(\vec r_{(P+1)}-\vec r_P)$. 
The rest of the polymer from P to its end is
translated in block by a vector $(\vec r'_{(P+1)}-\vec r_{(P+1)})$. 
Such moves are also combined with ``cluster moves'' of counterions
at distance less than $d$. Compared to pivots they have a
higher probability to be accepted and a suitable
combination of these both types of cluster-moves 
accelerates the convergence.

\begin{figure}
\resizebox{0.75\textwidth}{!}{%
  \includegraphics{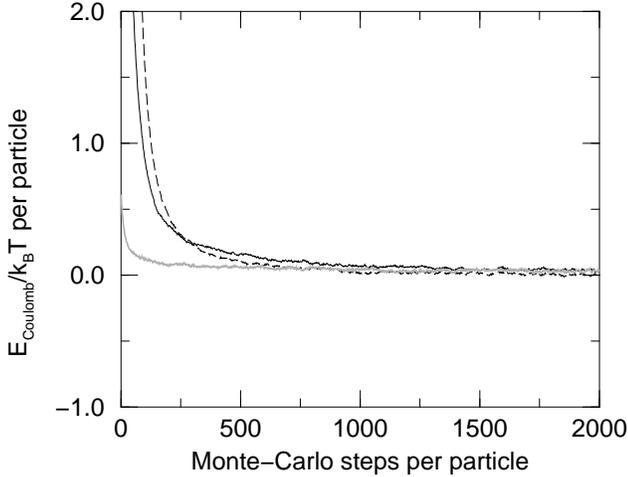}
}
\caption{Coulomb energy as a function of the number of steps at  $\zeta=2$ . 
Full black line: 
starting from fully stretched polymers. Dashed line: starting from the
configuration of neutral polymers. Full gray line: starting from 
the equilibrium
configuration at $\zeta=1$. Here $N=130$, $f=27$, $A=320a$. 
Beyond 1000 steps the differences between the three curves
are of the order of the fluctuations.}
\label{fig:1}       % Give a unique label
\end{figure}

\begin{figure}
\resizebox{0.75\textwidth}{!}{%
  \includegraphics{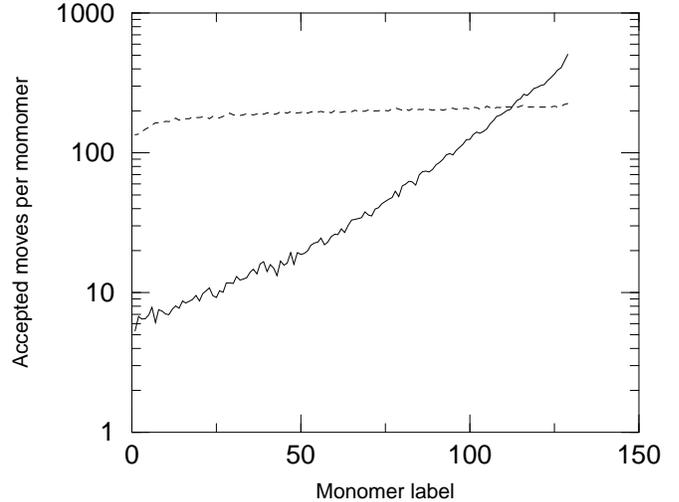}
}
\caption{Number of accepted ``cluster-moves'' per monomer as a function
of the monomer label (from 1 to $N$, starting from the core). 
Full line: ``clothed pivot rotations''; dashed line: 
``clothed translations'' (see text).}
\label{fig:2}       % Give a unique label
\end{figure}

Our main criterion for equilibrium is the convergence of the Coulomb
energy to the same stationary value starting from two drastically
different initial configurations:
\begin{itemize}
\item
Fully extended polymers with counterions distributed uniformly in the
volume of the cell.
\item
Contracted star with polymers in the equilibrium configuration
of uncharged polymer and counterions distributed uniformly in the volume
of the cell.
\end{itemize} 

%\noindent
As a typical example,
Fig. 1 illustrates the convergence of the Coulomb energy, starting from
these two extreme configurations for a micelle with 27 branches
of 130 monomers ($2\times 3510$ particles) at $\zeta =2$.
The equilibrium was reached after
$10^3$ steps per particle, representing 25 hours of monoprocessor
 time on a Compaq RS232 parallel computer.
The relevant statistical properties (distribution of monomers and
counterions, form factors etc...) were measured over $10^3$ 
further steps.
 A ``cluster move'' of a polymer part was attempted
every 3 iterations (the two others being devoted to single counterion
moves). The cluster moves were either pivots (3/4 of them) or 
translations (1/4 of them) as previously described. During the measure of
statistical averages,
each counterion was moved approximately
$ 1300$ times, about
$80$ pivots and $200$ translations per monomer were accepted.

The number of accepted pivots as a function of the position of
the pivoting point $P$ is represented in Fig.~2. Close to the core of
the micelle, the number of accepted moves is low (of the order of 10 per 
monomer). Nevertheless, it grows roughly exponentially as a function
of the position of monomer $P$ and statistics becomes acceptable
beyond a few tens of monomers.

The size $\delta x$ of the cubic box for elementary counterion 
displacements was
$10a$, and the elementary solid angle $\delta\Omega$ for pivot rotations
was fixed to $2\pi10^{-2}$ steradians. We have also tried some simulation
with  $\delta\Omega$ increasing with the distance to the core of the
micelle, without any substantial improvement of the rate of convergence
to equilibrium.

\section{General results}

\subsection{The theoretical model studied}

We  present here some theoretical results concerning 
the variations of the monomer and
counterion profiles as a function of the dimensionless parameter 
$\zeta$ [cf. equation (1)]
measuring the strength of the Coulomb interaction with respect to $k_BT$.

\vskip .3truecm
We consider two micelles with numbers of monomers and branches smaller than 
these corresponding to the experimental system studied in the next section:
\begin{description}
\item[(i)] $N=130$ and $f=6$
\item[(ii)] $N=130$ and $f=27$ 
\end{description}
The radius of the core $R_{core}$ is chosen to be small compared to the radius 
of the micelle corona and has thus a little influence on the corona.
We take in all cases $R_{core}=5a$. The influence of the boundary conditions
was investigated by comparing the results obtained with two different
sizes of the cubic cell: $A=320a$ and $A=640a$.

We  vary the dimensionless parameter $\zeta$ keeping the ratios  
$\tilde{A}=A/a$, $\tilde{R}_{core}=R_{core}/a$ and 
$\tilde{\sigma}_i=\sigma_i/a$ constant.
We start from a low value $\zeta_0=0.005$ and increase it by steps. At each
new step $\zeta_{n+1}$ we take as initial state of the Monte-Carlo
calculation the equilibrium state obtained at the preceding one
$\zeta_n$. The system is thus progressively ``annealed'', which provides
a faster convergence to equilibrium at each step. As a typical test for
equilibrium we compare in Fig.~1 the energy obtained in that way at
$\zeta=2$ starting from the equilibrium configuration at $\zeta=1$ to
the results obtained with far-from-equilibrium arbitrary initial
configurations. After 1000 iterations the differences between all
curves are of the order of the fluctuations.

We calculate the mean values of the radial distributions of monomers and
counterions.
The deviations of these distributions from the monomer distribution
corresponding to fully extended arms are
better emphasized by considering, in terms of
$\tilde{r}=r/a$, the dimensionless quantities:
\begin{equation}
\rho_{\{m,i\}}(\tilde{r})={ 4\pi\over f} 
\tilde{r}^2  a^3c_{\{m,i\}}(\tilde{r})
\end{equation}
where $c_{\{m,i\}}(\tilde{r})$ represent the local volume densities.

\noindent
{ For a micelle with fully extended arms, $\rho_m$ should be 1 for 
$\tilde{R}_{core}<\tilde{r}<\tilde{R}_{core}+N$ and 0 elsewhere.}

\subsection{Monomer profile as a function of $\zeta$}

For  $N=130$, $f=27$ and $\tilde{A}=320$,
typical monomer and counterion profiles at represented in Figs.~3a,
 3b and 3c at  weak coupling ($\zeta=0.02$), intermediate
coupling ($\zeta=0.2$) and strong coupling ($\zeta=3$) respectively.
Corresponding snapshots of the micelle are shown in Fig.~4.

The variations of the radius of gyration $R_g$ and of the corona 
radius $R_c$ as a function of $\zeta$ are shown in Fig.~5 (circles).
The corona radius $R_c$ is defined here as the abscissa of 
the maximum of the free-end distribution, 
which is approximately gaussian (see Fig.~6).
 It also roughly coincides with the rightmost inflexion point in the 
concentration profile $\rho_m(\tilde{r})$.

We have studied the influence of boundary conditions by doubling the size
$A$ of the cubic cell (crosses). $R_c$ and $R_g$ are very weakly dependent
on boundary conditions. Relative difference of the order of a few percent
at most are observed at maximum stretching ($\zeta$ of order 1).
We conclude that  it is not useful to investigate more sophisticated
treatments of Coulomb interaction like these corresponding to periodic
boundary conditions with Ewald sums.

The influence of the number of arms is seen in Fig.~5 by comparing our results
with a small number of arms $f=6$ (diamonds) to the results with
$f=27$ (circles). It is only important at weak coupling.

The Figs.~3, 4 and 5 show clearly, as a function of $\zeta$, 
three different regimes. These three regimes 
are  governed by changes in the behavior of the counterions, which are
expected from mean-field 
theories \cite{Borisov1,Borisov2,Borisov3,Manning,Luc1}.

\begin{figure}
\resizebox{0.75\textwidth}{!}{%
  \includegraphics{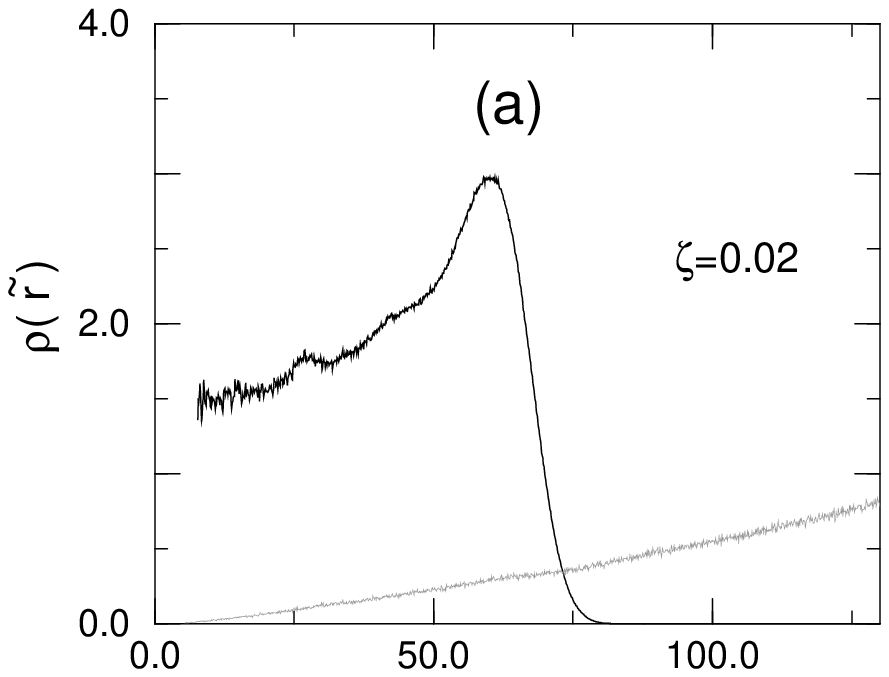}
}
\resizebox{0.75\textwidth}{!}{%
  \includegraphics{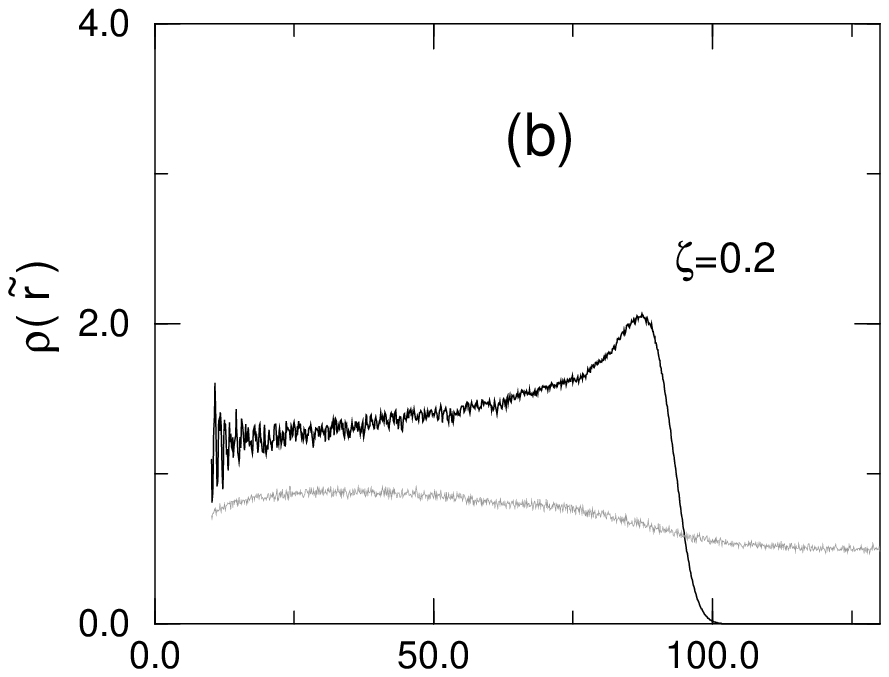}
}
\resizebox{0.75\textwidth}{!}{%
  \includegraphics{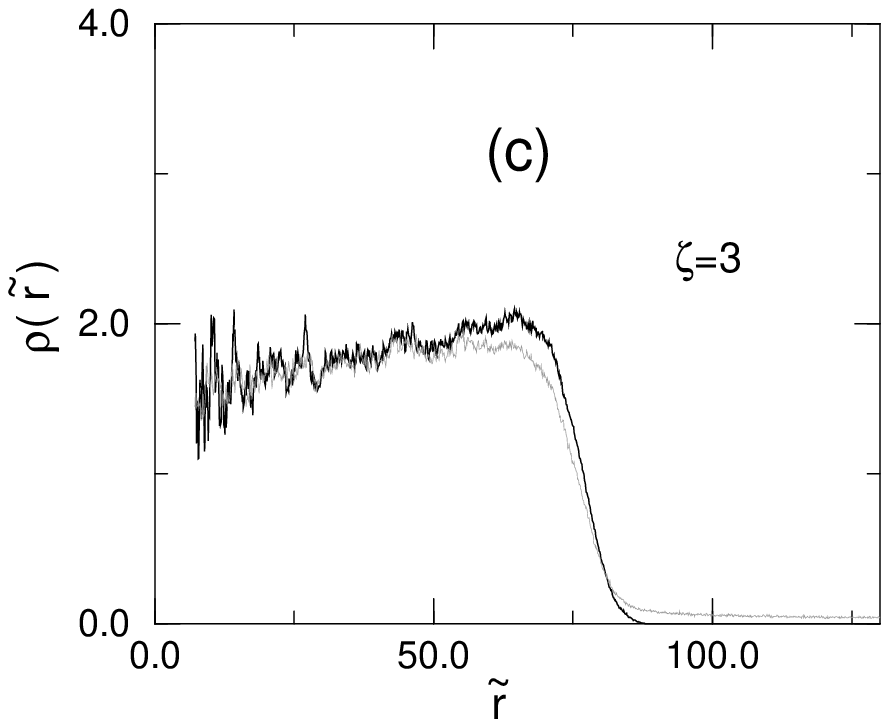}
}

\caption{Density profiles of monomers (black) and counterions
(grey) in the three different regimes: a) uncondensed
regime $(\zeta=0.02)$; b)  spherically condensed regime $(\zeta =0.2)$;
 c) Manning-condensed regime $(\zeta=3)$. Here $N=130$, $f=27$ 
and $\tilde{A}=320$.}
\end{figure}

\begin{figure}
\resizebox{0.75\textwidth}{!}{%
  \includegraphics{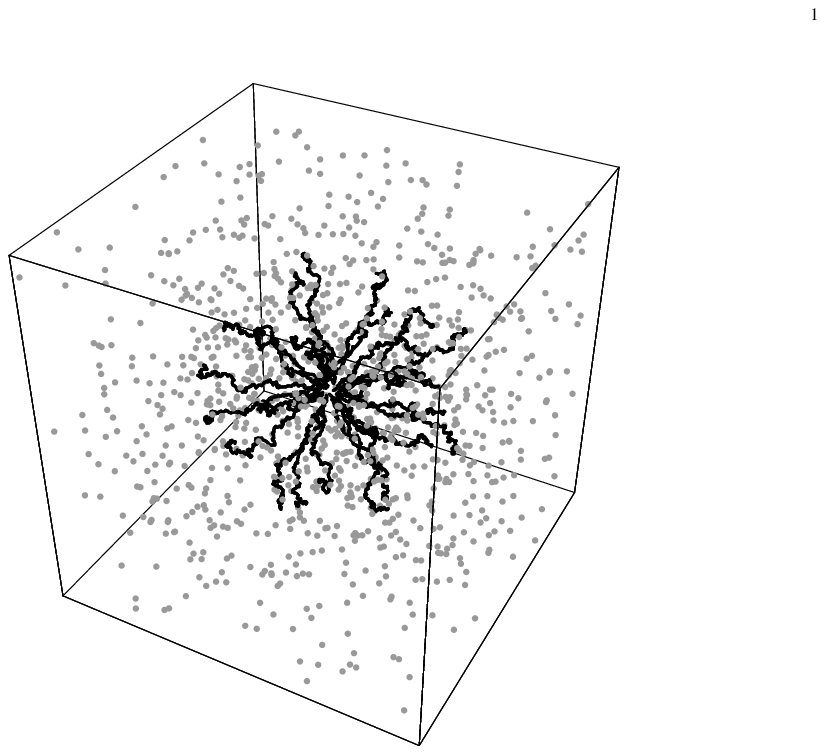}
}
\resizebox{0.75\textwidth}{!}{%
  \includegraphics{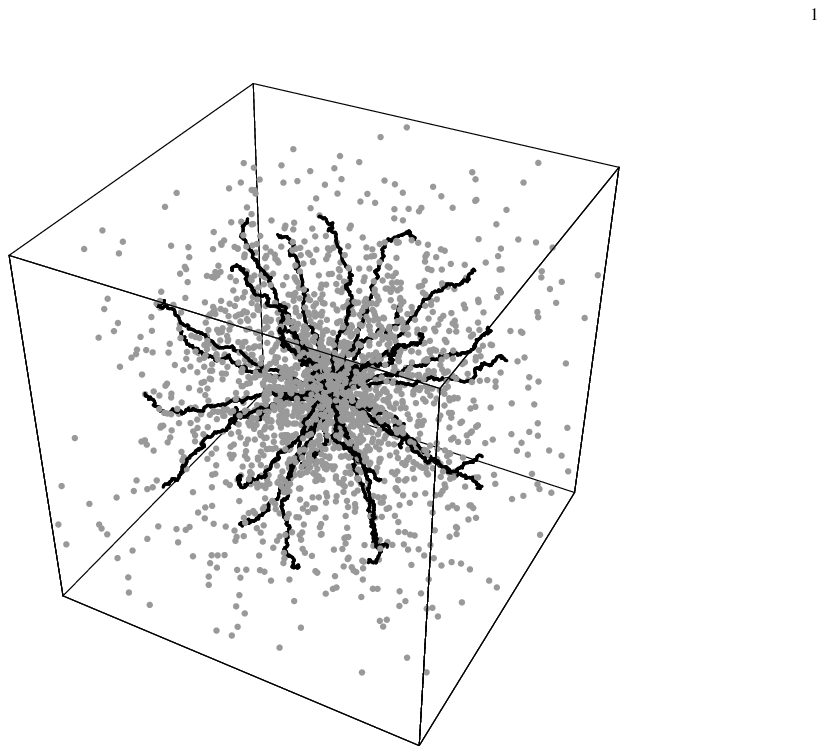}
}
\resizebox{0.75\textwidth}{!}{%
  \includegraphics{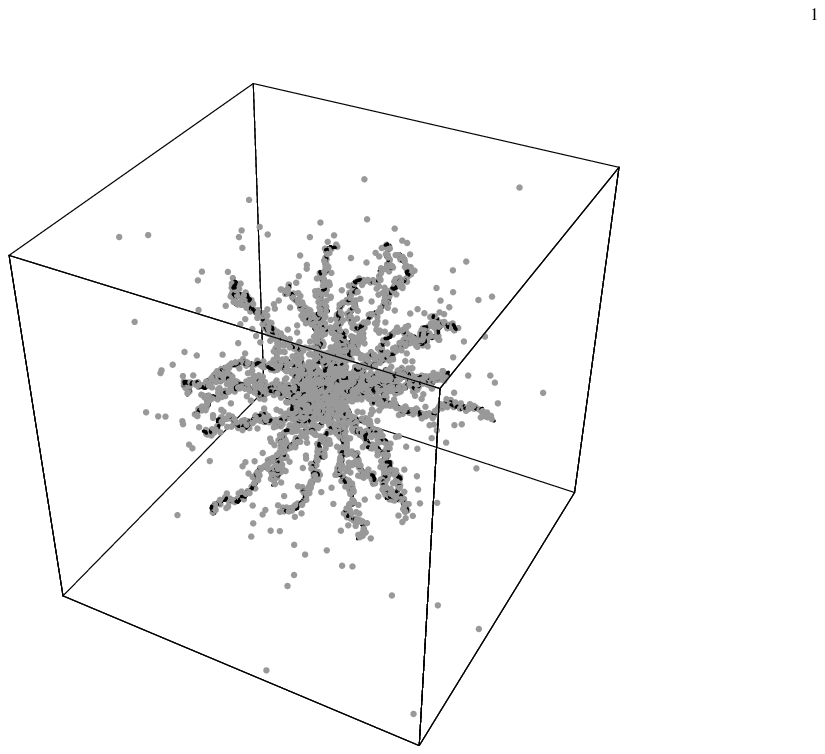}
}

\caption{Snapshots of the micelle in equilibrium corresponding 
to the three regimes considered in Fig. 3 respectively. 
The three snaphots are represented at the same scale.}
\end{figure}

\begin{figure}
\resizebox{0.75\textwidth}{!}{%
  \includegraphics{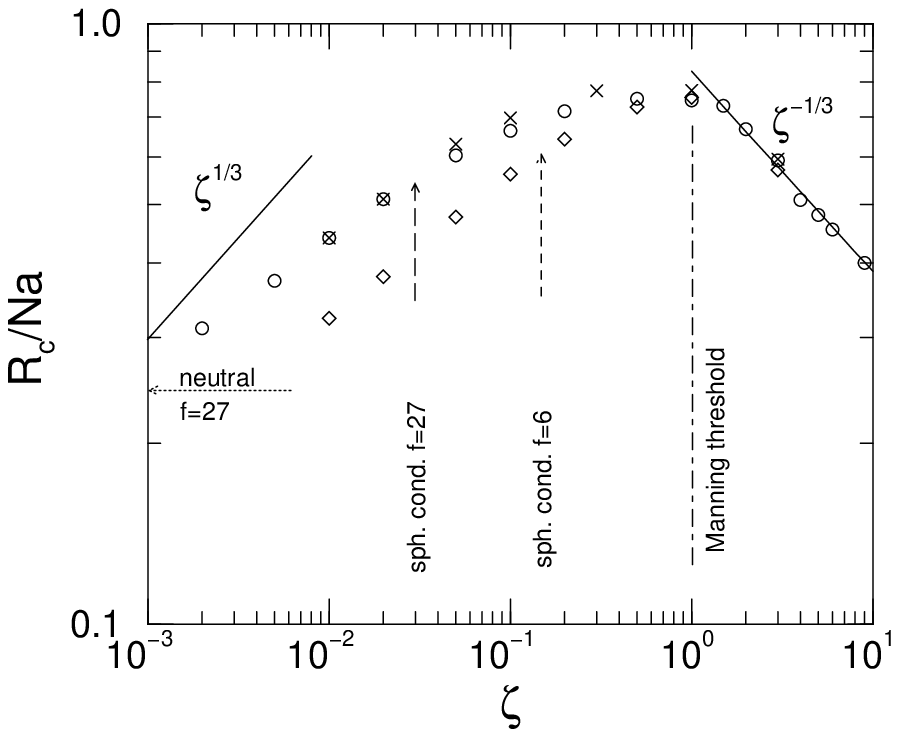}
}
\resizebox{0.75\textwidth}{!}{%
  \includegraphics{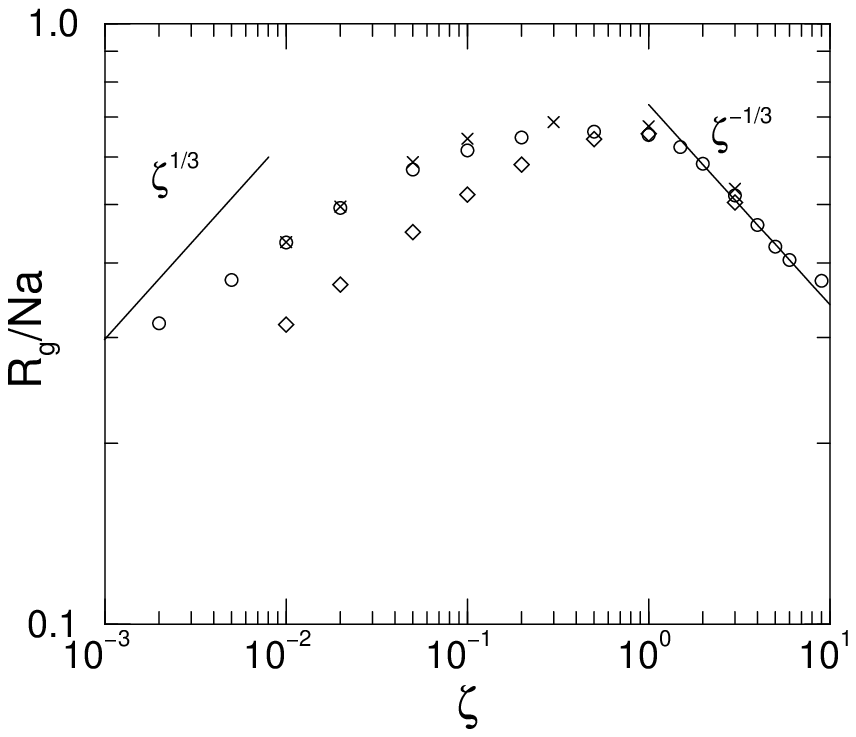}
}
\caption{Variation of the corona radius $R_c$ and  radius 
of gyration $R_g$ as
a function of the dimensionless parameter $\zeta$ for: (i)  $f=27$,
$\tilde{A}=320$ (circles); (ii)  $f=27$, $\tilde{A}=640$ (crosses); 
(iii) $f=6$,$\tilde{A}=320$ (diamonds). $N=130$ in all cases.
The Manning threshold is indicated by the dash-dotted vertical line
$\zeta\approx 1$.
The  dashed and dotted vertical lines indicate different
threshold values for spherical condensation corresponding to
(i), and (iii) respectively . The horizontal arrow on the left
indicate the limiting values of the corona radius corresponding to neutral 
stars with  $f=27$  arms of $N=130$ monomers.}
\end{figure}
 
\begin{figure}
\resizebox{0.75\textwidth}{!}{%
  \includegraphics{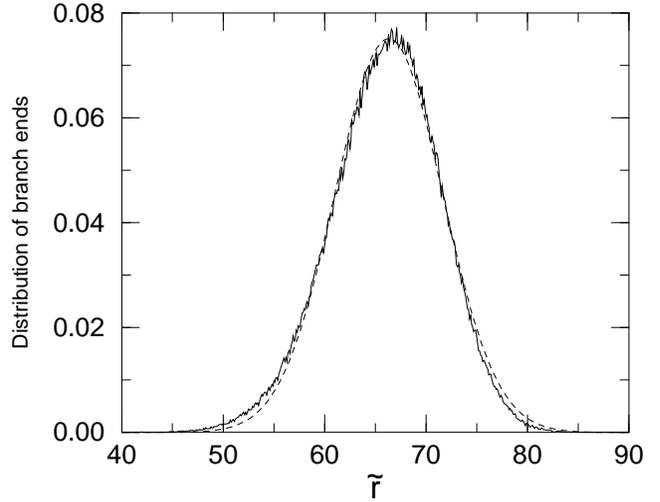}
}
\caption{Distribution of branch ends for $\zeta=0.02$. The Monte-Carlo results
(full line) are  fit by a gaussian (dashed line):
$1/[\sqrt{2\pi}(\tilde{\sigma})]
\exp [-(\tilde{r}-\tilde{R}_c)^2/2\tilde{\sigma}^2]$ 
with $\tilde{R}_c=66.33$ and
$\tilde{\sigma} =5.31$.}
\label{fig:6}       % Give a unique label
\end{figure}

\begin{figure}
\resizebox{0.75\textwidth}{!}{%
  \includegraphics{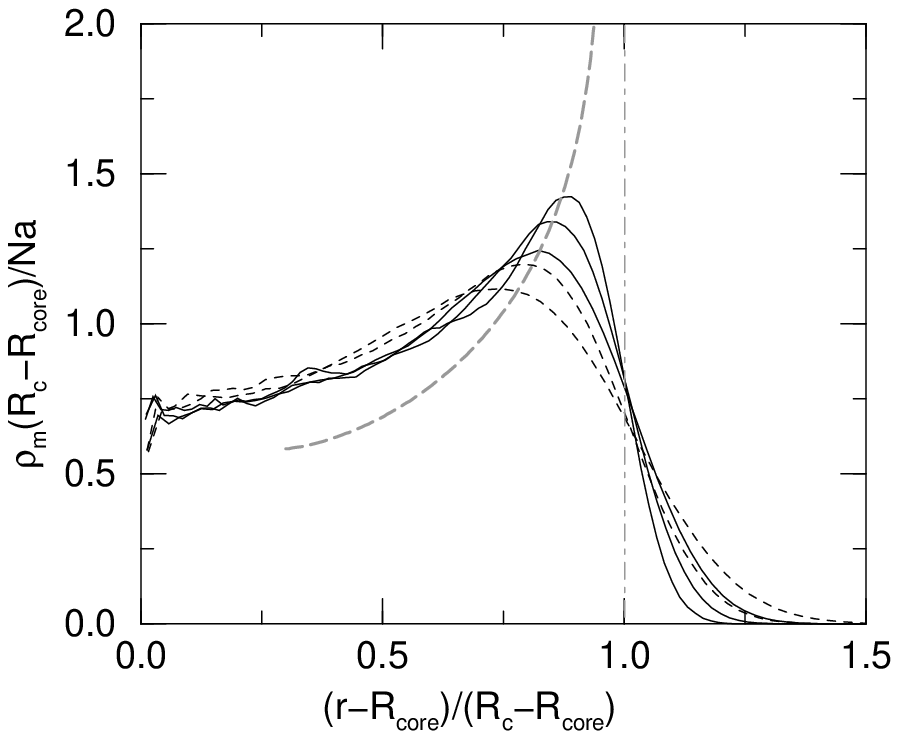}
}
\caption{
Monomer profiles in the ``unscreened'' regime. Full lines: $N=130$ and $f=27$
for three $\zeta$-values, $\zeta=0.005$ (lowest maximum), $\zeta=0.01$ and
$\zeta=0.02$ (highest maximum). Dotted lines:  $N=130$ and $f=6$ for 
$\zeta=0.01$ and $\zeta=0.02$. The gray dashed line indicates the solution of
Eq.~7.
}
\label{fig:7}       % Give a unique label
\end{figure}

\subsection{The ``unscreened regime''.}

At weak coupling $\zeta<<1$, the contribution of the counterions to the
entropy dominates  the electrostatic energy. The local volume 
density $c_i$ of counterions is roughly constant in the cell:
$c_i\approx fN/A^3$ and from equation (4)
$\rho_i\approx 4\pi N \tilde{r}^2/\tilde{A}^3$. 
If $A$ is larger than $R_c$, the fraction of counterions 
in the micelle corona is small and
the electrostatic interactions between branches are  weakly screened.

In the unscreened limit, the self-consistent field arguments proposed
a long time ago for linear polyelectrolytes by de Gennes et al. \cite{PGDG} 
have been straightforwardly generalized to star-branched 
polyelectrolytes \cite{Borisov1,Borisov2}. Let us consider a chain
whose first unit is placed at the origin. The $n$-th unit is at an average
distance $r(n)$ and feels two forces:
\begin{itemize}
\item
In a simple gaussian model, there is an elastic force which tends to contract
the chain:
\begin{equation}
F_e=
%{3k_BT\over a^2}[({\vec r}_{n+1}-{\vec r}_{n})-({\vec r}_{n}-{\vec r}_{n-1})]=
{3k_BT\over a^2}{d^2r\over dn^2}.
\end{equation}
\item
If we assume equal stretching of the $f$ chains of a star, 
{\it i.e.} all chain ends are at the same distance $R_c$ from the origin,
the spherical region at distance smaller than $r(n)$ contains a mean
charge $Q=nfz_me$. Applying Gauss theorem, there is an electrostatic force:
\begin{equation}
F_c={f(z_me)^2\over 4\pi\epsilon\epsilon_0}{1\over r(n)^2}.
\end{equation}
\end{itemize}
Writing the equilibrium between the two forces, one obtains  the following
dimensionless equation:
\begin{equation}
{d^2u\over dx^2} +{1\over 3}{x\over u^2}=0 
\end{equation}
with $x=n/N$ and $u= r/(Nd)$ where the length scale $d$ is:
\begin{equation}
d=a\left(f z_m^2l_B\over a\right)^{1/3}
\end{equation}
The boundary conditions are $u(0)=0$ and $du/dx=0$ at $x=1$.
Taking $f=1$ in the previous algebra is nothing else than De Gennes et
al. equations \cite{PGDG}.

The center to end distance of the chains is thus proportional to $Nd$:
\begin{equation}
R_c\approx f^{1/3}\zeta^{1/3}Na
\end{equation}
(we take $z_i=-z_m$).

These simple scaling arguments lead to a stretching of the chains which
is proportional to the number $N$ of monomers. 
They have justified a so called ``rod-like'' picture of the branches.

In Fig.~5, the slope $d\log {R_c}/d\log {\zeta}$ is smaller than $1/3$,
the value expected through the previous scaling arguments. 
The dependence of $R_c$ on $f$ is also weaker than predicted
by equation (9). At very weak coupling, we obtain (see Fig.~5) 
$R_c(f=27)/R_c(f=6)\approx 1.35$ compared to the expected scaling:
$(27/6)^{1/3}=1.65$. The explanation is that  when $\zeta$ is 
low enough to ensure that very few counterions are
inside the corona, the excluded volume
effects (which are not taken into account in the previous scaling arguments) 
become significant. The value of $R_c$ has a lower limit which is given
by the scaling law for neutral stars \cite{Daoud}:
\begin{equation}
R_c^{neutral}\approx a f^{1/5}N^{3/5}
\end{equation}
This limiting value for $N=130$ and $f=27$, within our pearl-necklace model is
indicated in Fig.~5. Our lowest value for $R_c$ shows an inflexion 
towards this asymptote.

Close to the core, our monomer profiles can be roughly scaled on a universal
curve by plotting $\rho_m(R_c-R_{core})/Na$ as a function of 
$(r-R_{core})/(R_c-R_{core})$ (i.e.  the
integral of each curve is normalized to 1) and compared to the solution of
Eq. 7 given in Fig. 2 of Ref. \cite{Borisov1} (see Fig.~7). 
Close to the core, our Monte-Carlo results agree qualitatively with 
the  solution of equation (7) which gives a roughly constant profile of
$\rho_m$ (i.e. 
$a^3c_m(\tilde{r})\approx \tilde{r}^{-2})$. 
At $R_c$, equation (7) leads to some unphysical divergence, which 
is smoothed in  more refined SCF models taking into account the fluctuations
of the chain ends \cite{Borisov1,Borisov3}. Nevertheless a substantial
increase of $\rho_m$ near $R_c$ remains in these SCF models, as well as in
our simulation.

\begin{figure}
\resizebox{0.75\textwidth}{!}{%
  \includegraphics{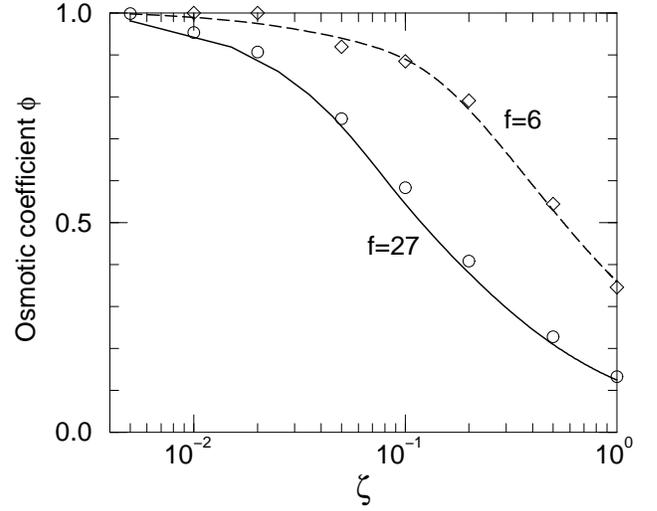}
}
\caption{
Osmotic coefficient $\phi$.
The open diamonds and circles represent the Monte-Carlo results for $f=6$
and $f=27$ respectively. The full and dashed line correspond to the
Poisson-Boltzmann approximation. In both cases $\tilde{A}=320$ and $N=130$. 
}
\label{fig:8}       % Give a unique label
\end{figure}

\begin{figure}
\resizebox{0.75\textwidth}{!}{%
  \includegraphics{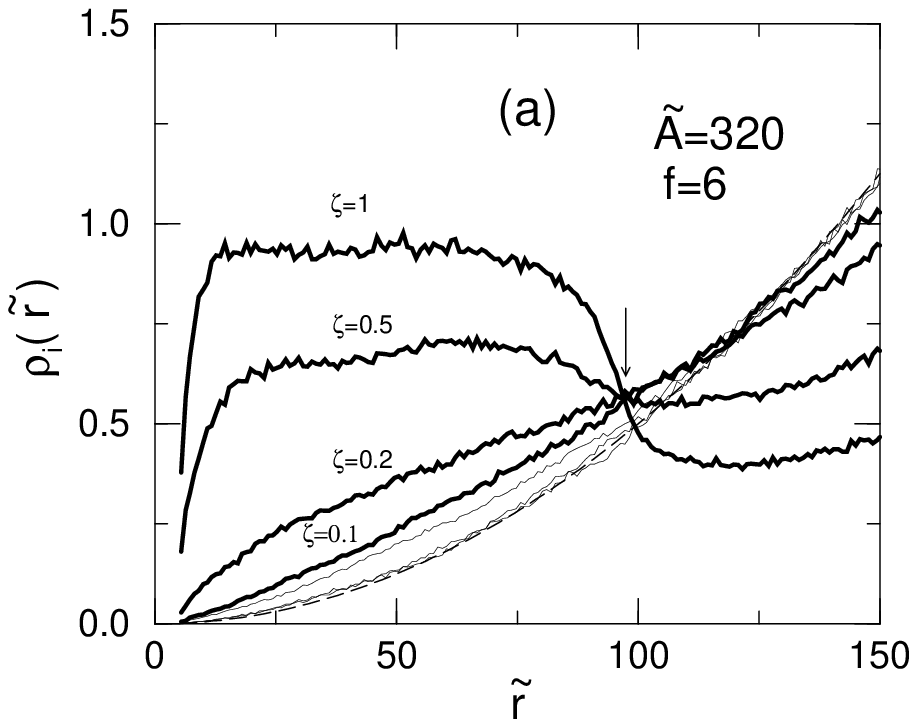}
}
\resizebox{0.75\textwidth}{!}{%
  \includegraphics{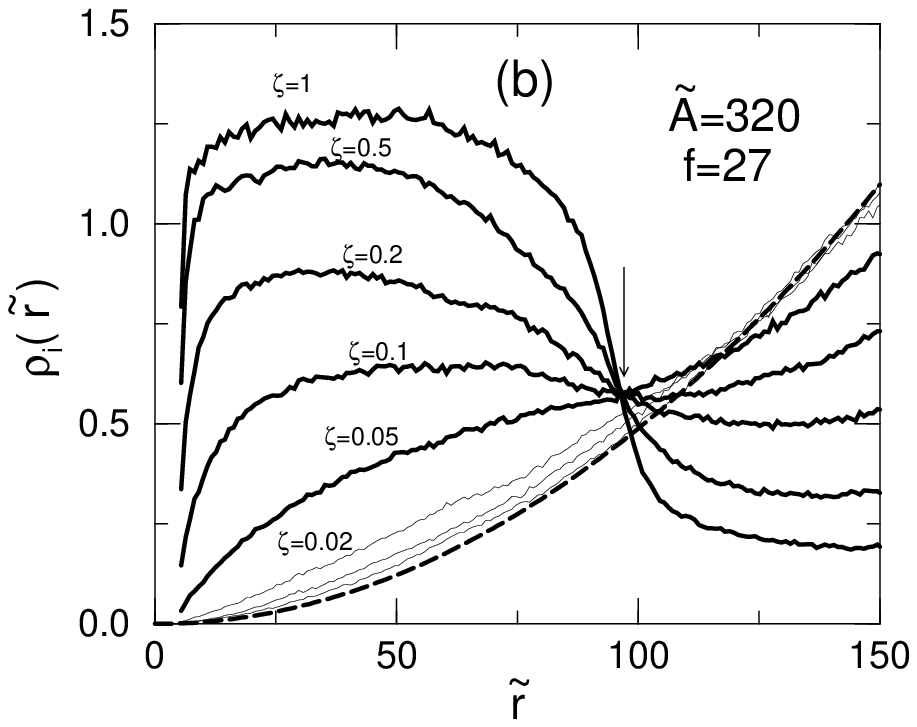}
}
\resizebox{0.75\textwidth}{!}{%
  \includegraphics{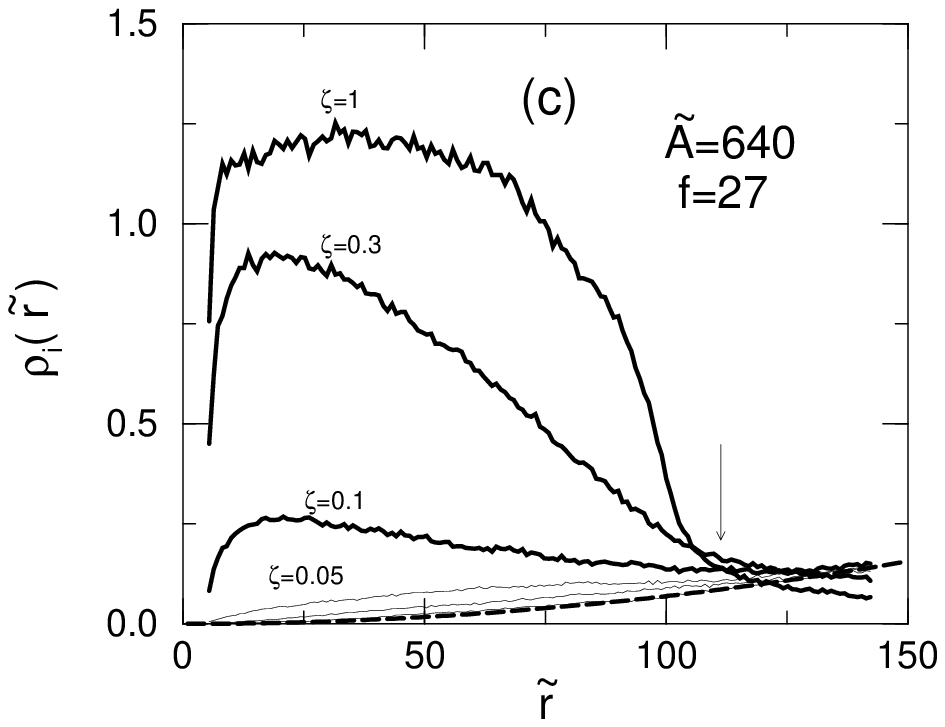}
}

\caption{Radial distribution of counterions (N=130) as a function 
of $\zeta$ for
different numbers of arms: (a) $f=6$ and $A=320$, (b) 
 $f=27$ and $A=320$, and different cell sizes: 
(c)  $f=27$ and $A=640$. The arrows indicate a common inflexion
point at $r\approx R_c$ in the pure spherically-condensed regime. The dashed
parabola indicate the equi-partition of all counterions in the volume of
the cell in the uncondensed regime.}
\label{radcond}
\end{figure}

\subsection{The ``osmotic regime'' with spherical 
condensation of counterions}

Increasing $\zeta$ we expect a ``spherical condensation'' of a part of the
counterions inside  and in the vicinity of the corona. This occurs 
when the Coulomb energy of a counterion in
the field created by the charged monomers: 
$$
(z_mNf)z_ie^2/(4\pi\epsilon\epsilon_0 R_c)
$$
at the limit of the corona radius is of the order of $kT$, i.e. for
$$
\zeta_{sph} \approx {1\over f}
$$
(at this level of approximation we take $R_c\approx Na$).
The condensation of ions near the surface of an impenetrable 
charged sphere \cite{Luc1} has been formulated
in an analogous way to the cylindrical 
condensation of ions around a charged 
rod studied by Manning \cite{Manning}.
 The  more complex problem of counterion condensation in a penetrable
micelle has been studied through Self-Consistent Field 
methods \cite{Borisov1,Borisov2,Borisov3}. It can also be formulated
in a simpler way through a Poisson-Boltzmann analysis.
The
Poisson-Boltzmann equation can be solved numerically for a penetrable sphere
of radius $R_c$ with charge density 
$$c_m(r)=fN/(4\pi R_cr^2)$$
inside a spherical cell of volume $A^3$.
The Fig. 8 compares the Monte-Carlo to the Poisson-Boltzmann results for 
the osmotic coefficient $\phi$. The osmotic
coefficient is defined as the ratio of the osmotic pressure to its value
corresponding to a uniform repartition of the ions in the cell. The osmotic
pressure is equal to $k_BT$ times the local ionic density at the edge
of the cell.

There
is a clear change of slope of $\phi(\zeta)$ at a point which can be 
defined as the crossover to the condensed regime.
For different numbers of arms and cell sizes, we deduce
the crossover to the spherically condensed regime at the following values
$\zeta_{sph}$:
\begin{itemize}
\item
 $\zeta_{sph}\approx 0.15$ for  $f=6$, $\tilde{A}=320$ 
\item
 $\zeta_{sph}\approx 0.03$ for  $f=27$, $\tilde{A}=320$ 
\item
 $\zeta_{sph}\approx 0.08$ for  $f=27$, $\tilde{A}=640$ 
\end{itemize}
($N=130$ in the three cases). At fixed $\tilde{A}$, $\zeta_{sph}$ varies in
$1/f$.

In Fig.~9, we show that at the thresholds defined above, there is a drastic
change of the counterion profile.  Below $\zeta_{sph}$, 
the local density $c_i$ of counterions
is roughly constant $c_i\approx fN/A^3$ leading to
a quadratic behavior of $\rho_i$. Above the threshold, the curves $\rho_i$
show an inflexion point at $\tilde{r} \approx R_c/a$. Since in that 
$\zeta$ range
$R_c$ is roughly constant, the abscissa of this inflexion point remains
constant. 

For $\zeta>>1/f$, most of the counterions are inside the micelle 
corona (see Fig.~9).
This regime has often been called ``osmotic'' by reference to the
osmotic pressure of counterions inside the corona \cite{Borisov2,Borisov3}.

The range $1/f<<\zeta<1$ corresponds to the largest stretching of the star
with a flat maximum in the curve $R_c(\zeta)$ (see Fig.~5).

\subsection{The ``Manning condensed'' regime}

\begin{figure}
\resizebox{0.75\textwidth}{!}{%
  \includegraphics{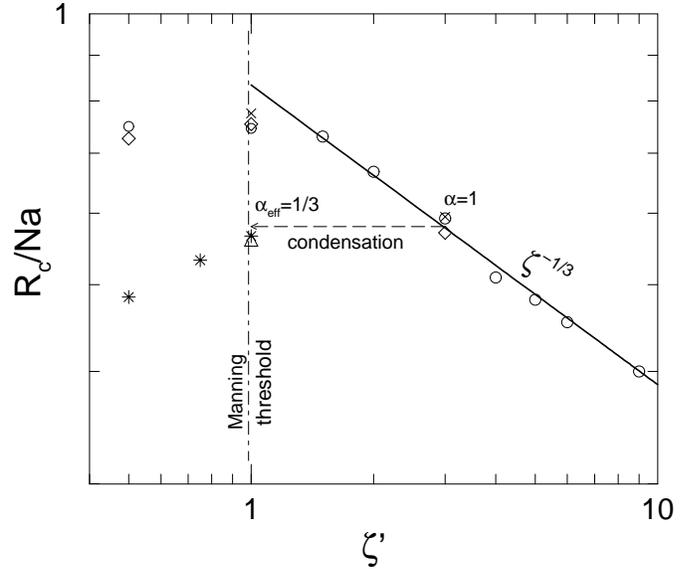}
}
\caption{ Variation of the corona radius near and in the Manning
condensed regime. The
stars represents the results of JLL [7] for $N=50$ and $f=18$ at charge
fraction $\alpha=1/6,~1/4$, 1/3, the triangle
corresponds to their results for $N=150$, $f=10$ and $\alpha=1/3$, the
 abscissa is $\zeta'=\alpha \zeta$. 
In JLL [7], all simulation are at constant $\zeta=3$.
The other symbols are
our results reported in Fig.~5  with the same conventions
(in our study $\alpha$ is fixed to 1 and $\zeta=\zeta'$ is varied). 
The dashed arrows indicates
a mapping of our model at $\zeta=3$ and $\alpha=1$, in the strongly 
condensed Manning regime, onto an effective condensed system with 
charge fraction $\alpha_{eff}=1/3$. It agrees nicely with the real system at
$\alpha=1/3$ studied by JLL at the same $\zeta=3$. }
\end{figure}

For larger $\zeta$ values, beyond 
the ``radial condensation'' of counterions in 
the micelle corona, we 
expect  the superposition of a ``cylindrical condensation'' of counterions 
around each branch of
the micelle. This phenomenon is clearly seen in Fig.~3c and Fig.~4c.
 The Manning criterion for this ``cylindrical condensation''
is \cite{Manning,Luc1}:
$$
\xi_M= 1/|z_i|
$$
which corresponds to the threshold value $\zeta_M=1$. 

The Fig.~10 represents an enlargement of the last decade in Fig.~5. 
The corona radius $R_c$ presents a flat maximum at $0.5<\zeta<1$ and
decreases at $\zeta>1$.
At this stage, it is interesting to compare our approach with the
Molecular Dynamics study by JLL \cite{Likos1,Likos2} where $\zeta$ has
been fixed to 3, but where only one monomer out of three, four or six is
charged. The ``charge fraction'' $\alpha$ is thus varied from 
$\alpha=1/3$ to $\alpha=1/6$ whereas it is $\alpha=1$ in our model.
The relevant Manning parameter which characterizes
the linear charge density of the polymer is $\xi=z_ml_B/b$, 
where $b=a/\alpha $
is the distance between consecutive charged monomers \cite{Footnote1}. 
In Fig.~10 we have
plotted some of JLL results, as a function of $\zeta'=\alpha \zeta$.
The fact that from $\alpha =1/6$ to $\alpha=1/3$ their corona radius
increases indicates that they have not reached the Manning condensed
regime at $\alpha=1/6$ and $\alpha=1/4$. They are just reaching it 
near $\alpha=1/3$.

\begin{figure}
\resizebox{0.75\textwidth}{!}{%
  \includegraphics{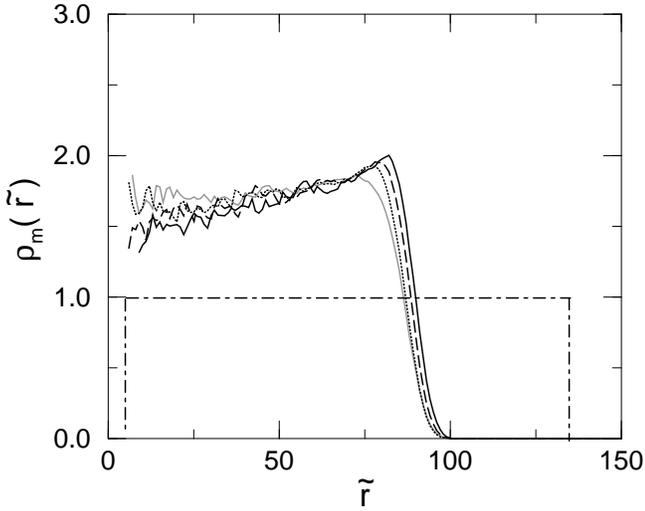}
}
\caption{Monomer density profiles for $\zeta=2$,
 $N=130$ and $f=6$ (dotted line), $f=27$ (dashed line),  
and $f=54$ (full black line). They are compared to the density profile
of one branch attached to a fixed center point with $R_c=0$ (grey line).
Since $R_c=5a$ for all other curves this last grey line has been translated
by 5 units in $x$ direction. The dot-dashed line indicate the profile
corresponding to fully extended branches.
Only a slight increase of the corona radius is observed when the number of
branches is varied by a factor 10 (from $f=6$ to $f=54$)} 
\end{figure}

\begin{figure}
\resizebox{0.75\textwidth}{!}{%
  \includegraphics{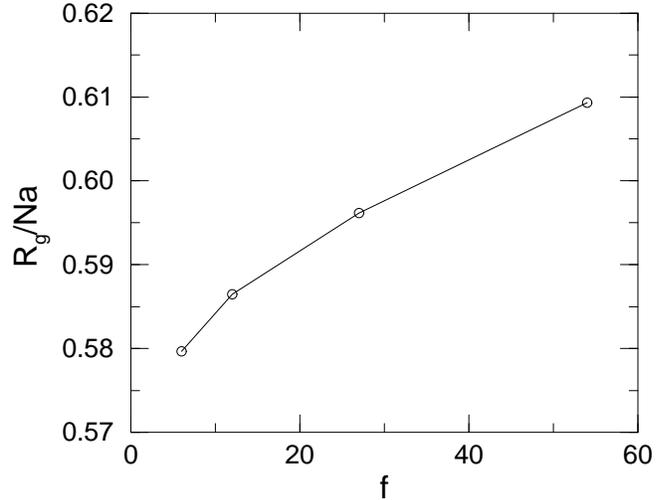}
}
\caption{Variation of the gyration radius $R_g$ as a function of 
the number of branches $f$ for $\zeta=2$ and
 $N=130$.
Only an increase of a few percent of $R_g$ is observed when the number of
branches is varied by a factor 10} 
\end{figure}

At $\zeta>> 1/f$, (near the broad maximum of $R_c$ and in  the Manning
condensed regime), the monomer profiles are only very 
weakly dependent on $f$. 
In Fig.~11, we compare the monomer 
profiles  for different values of the number of branches
$f=6$, $27$, and $54$.
 The number of monomer per branch is
$N=130$, the radius of the hydrophobic core is $\tilde{R}_{core}=5$ 
and the size of the cell is $\tilde{A}=320$.
The Fig.~12 represents the variation of the gyration radius as a function
of the number of branches $f$. It increases only by  a few percent when
the number of branches vary by a factor of  10. 
The MD results of JLL (see Tables I and II in Ref. \cite{Likos2}),
also show a weak dependence of $R_c$ on $f$.

\noindent
The Fig.~13 represents a blow up over the last 80 monomers of one branch
in Fig.~4c. 
At a length scale of the order of 10 monomers, a local ``rod-like'' picture
is relevant.
To illustrate the phase space explored by a given branch, we have registered
 100 snapshots of the same branch (Fig.~14), at equal time intervals
in thermodynamic equilibrium.
The phase space explored corresponds roughly to a cone of solid angle
$4\pi/f$ steradians which shows that each branch occupies a fraction $1/f$ 
of the sphere and rarely ``mixes''
 with neighboring branches. An ``urchin-like''
picture \cite{Muller2} is thus relevant.
 
\vskip .5truecm
The last decade in Fig.~5 and its enlargement in Fig.~10 show that, in our
fully charged model ($\alpha=1$), beyond the Manning threshold, the corona
radius $R_c$ decreases in $\zeta^{-1/3}$. This is  quite a new result.
Both this exponent and the independence of $R_c$ on $f$ can be interpreted
with the following simple arguments. 

  For $\zeta>1$, there is a
cylindrical condensation of a fraction  of counterions
along the branches. The  model proposed by Manning \cite{Manning},
considers that a fraction $(1-1/\zeta)$ 
of the counterions condenses around the branches 
in such a way that the monomers get an
effective charge $\alpha_{eff}z_m$ with an effective charge fraction
$$
 \alpha_{eff}=1/\zeta .
$$
The remaining fraction $1/\zeta$ of counterions are only spherically 
condensed and distributed between the branches. We assume that $f$ is
sufficiently high ($\zeta>>1/f$) to neglect the fraction of ``free''
counterions outside the corona.
Taking a spherical distribution in $1/r^2$, we obtain the following
effective ion density:
$$
c_i^{eff}(\tilde{r})={1\over \zeta}{ fN\over 4\pi a^2 R_c \tilde{r}^2}.
$$
The corresponding ``effective'' screening length is
$$
\kappa_{eff}^{-1}(\tilde{r})=[4\pi l_B z_i^2 c_i^{eff}]^{1/2}
=a\tilde{r}\left[f{|z_i|\over|z_m|}{Na\over R_c}\right]^{-1/2}.
$$
Since we restrict our study to $|z_i|=|z_m|$, it is
independent of $\zeta$. The ``effective'' screening length 
roughly keeps the
same value it reaches at the Manning condensation threshold $\zeta=1$.
At a radial position $r$, the mean transverse distance between two
neighboring branches is roughly $4a\tilde{r}f^{-1/2}$. It exceeds the
screening length by a factor larger than 4. Hence we can neglect the
electrostatic interactions between branches. Since the polymers are
relatively extended, the excluded volume interactions between branches
are also weak. This is the reason why the monomer profiles are so weakly
dependent on $f$.

\begin{figure}
\resizebox{0.75\textwidth}{!}{%
  \includegraphics{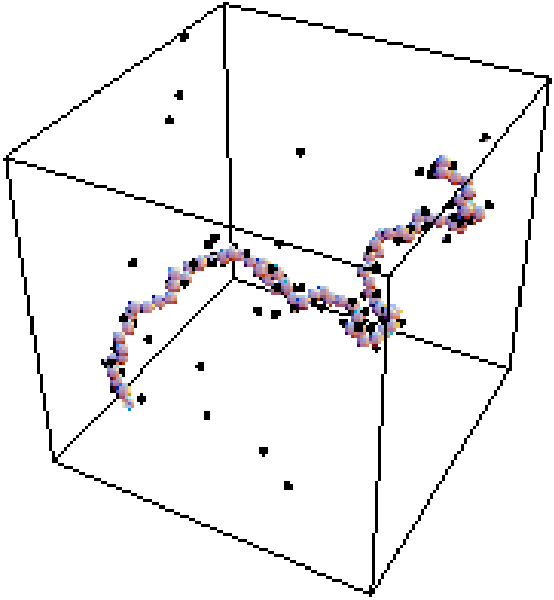}

}
\caption{
A blow-up over the last 80 monomers of one branch in Fig.~3c. Large spheres
are fully charged monomer. Small spheres with diameter 1/2 of the large
ones are counterions
} 
\end{figure}

\begin{figure}
\resizebox{0.75\textwidth}{!}{%
  \includegraphics{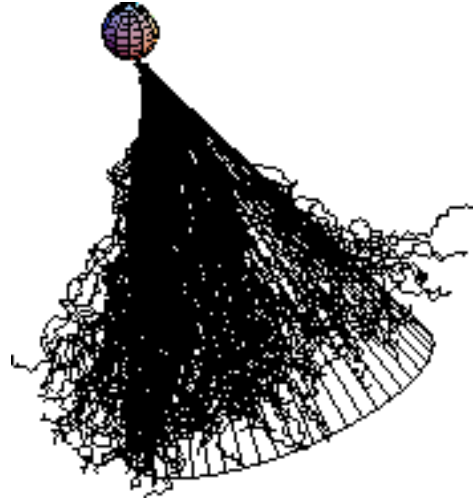}
}
\caption{Snapshots of the same branch at equal time intervals during
measurement of thermodynamic quantities. The portion of space explored
corresponds to a cone of solid angle $4\pi/f$ steradians. 
The sphere at the top
represents the hydrophobic core.
} 
\end{figure}

\vskip .3truecm

To justify the $(-1/3)$ exponent, we consider one branch in the
familiar electrostatic ``blob'' picture\cite{PGDG}.
Let us represent a branch as a succession 
of $N/m$ blobs of size $l$, each
containing $m$ monomers. As far as the size of a blob is smaller than
the screening length $\kappa_{eff}^{-1}$, 
the electrostatic energy of a blob is 
$$E_c=(\alpha_{eff}z_me m)^2/(4\pi\epsilon\epsilon_0 l).$$ 
Its elastic energy from entropic origin
is $$E_e=(3/2) k_BTl^2/(m a^2).$$ 
Minimizing the sum $E_c+E_e$ with respect to
$m$, we obtain:
$$
(l/m a)= \left[{4\over 3}
{(\alpha^{eff})^2z_m^2 l_B\over a}\right]^{1/3}
=\left[{4\over3}{|z_m|\over |z_i|}\zeta\right]^{-1/3}
$$
and the length of the branch is
$$R_c=(N/m) l=Na \left[{4\over 3}{|z_m|\over |z_i|}\zeta\right]^{-1/3}.$$ 
Since in the present study $|z_m|=|z_i|$, $R_c$ is proportional
to $\zeta^{-1/3}$ \cite{Footnote2}.

At  high $\zeta$, Schiessel and Pincus \cite{Pincus2} predicted a
collapse of the polymers with the following arguments. When the 
Coulomb energy dominates over the entropic contribution,
each counterion is practically fixed to a monomer and forms an electric
dipole. The attraction of dipoles should lead to a transition towards
 a collapse
of the star. Up to $\zeta=10$ our results follow  a $\zeta^{-1/3}$
power law, and we do not observe any precursor of this effect.

\vskip .5truecm
The Manning picture allows us to make some approximate mapping of our
model at $\alpha=1$, $\zeta=3$ onto the model of JLL at $\alpha=1/3$ and
the same $\zeta=3$. In our model, at $\zeta=3$, due to Manning condensation,
$(1-1/\zeta)=2/3$ of the counterions condense on the monomers. The remaining
effective charge fraction of monomers is thus $\alpha_{eff}=1/3$ 
and the other one third of the counterions
is only spherically condensed inside the corona. This effective model is
thus close to the real situation studied by JLL at $\alpha=1/3$ and
$\zeta=3$. The arrow in Fig.~10 shows a very nice agreement of the
corona radii corresponding to this mapping. This is a strong evidence
of the relevance of Manning's picture.

\vskip .5truecm

Up to now, radial distributions of counterions were only considered.
In the Manning regime, it is of prime interest to study the cylindrical
distribution of counterions around one branch.
Defining the distance $d$ of a counterion to the closest branch
as the distance to the closest monomer  we have measured the
mean value of the radial density $n (r,d)$, $n (r,d)\delta r\delta d$
representing the number of counterions at distance between $d$ and 
$(d+\delta d)$ from a branch and at distance between $r$ and $(r+\delta r)$
from the center of the micelle.

\begin{figure}
\resizebox{0.75\textwidth}{!}{%
  \includegraphics{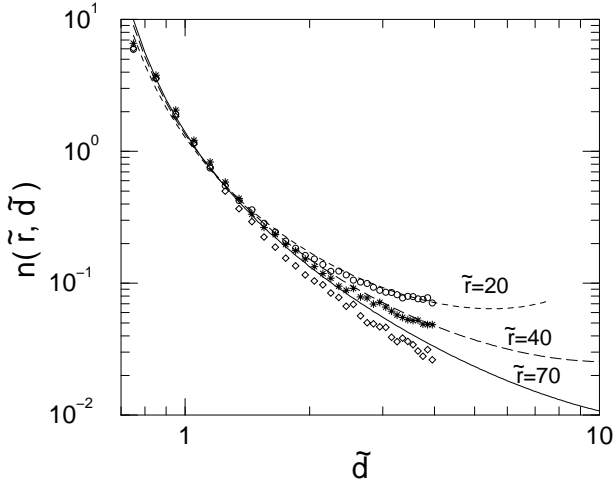}
}
\caption{Density of counterions at a radial distance $\tilde{r}$
 from the micelle
center and a transverse distance from the nearest branch $\tilde{d}$. 
The results
of the Monte-Carlo simulation (symbols) are compared to those of a
Poisson-Boltzmann approximation (continuous lines). (a) Circles and
dotted line: $\tilde{r}=20$; (b) stars and dashed line: 
$\tilde{r}=40$; (c) diamonds
and full line: $\tilde{r}=70$. Here, $\zeta=3$, $N=130$, $f=27$ 
and $\tilde{A}=320$.
} 
\end{figure}

\begin{figure}
\resizebox{0.75\textwidth}{!}{%
  \includegraphics{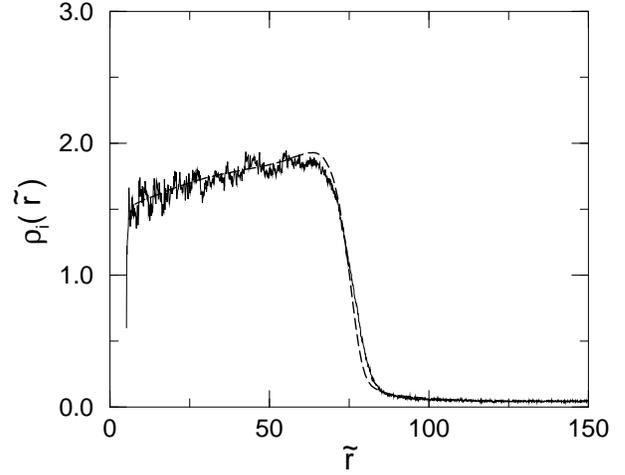}
}
\caption{
Radial density of counterions. The Monte-Carlo results are compared
to the Poisson-Boltzmann approximation (dashed line).
Here, $\zeta=3$, $N=130$, $f=27$ and $\tilde{A}=320$. 
} 
\end{figure}

\begin{figure}
\resizebox{0.75\textwidth}{!}{%
  \includegraphics{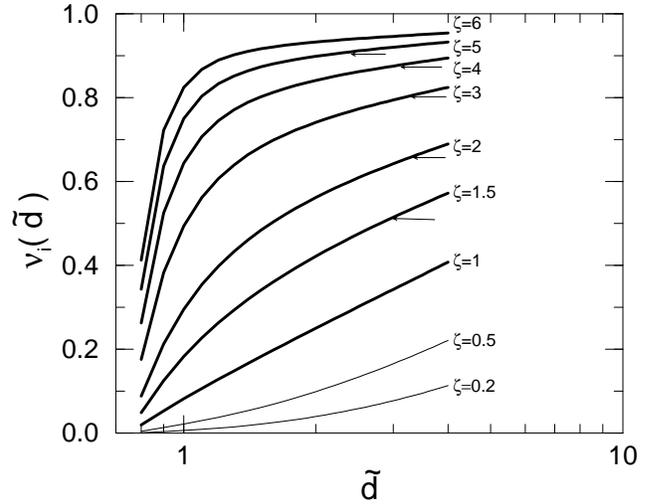}
}
\caption{ Number $\nu_i(\tilde{d})$  of counterion per monomer at distance less
than $d$ from
the closest branch, for different values of the dimensionless
parameter $\zeta$. Here $N=130$, $f=27$ and $\tilde{A}=320$. 
The change of curvature 
between $\zeta =0.5$ and $\zeta =1.5$ corresponds to Manning condensation.
The Manning threshold $\xi =1/z_i$ 
is reached at $\zeta \approx 1$.
A logarithmic dependence of $\nu$ on $\tilde{d}$ is expected 
at $\xi =1/z_i$, from
a Poisson-Boltzmann approximation in a ``rod-like'' picture. For each value
of $\zeta$ in the Manning condensed regime, the arrow indicates the distance
at which the ``effective charge parameter'' $\alpha_{eff} =1-\nu_i$ leads to
$\xi_{eff}=1/z_i$. }
\label{fig:4}       % Give a unique label
\end{figure}

\begin{figure}
\resizebox{0.75\textwidth}{!}{%
  \includegraphics{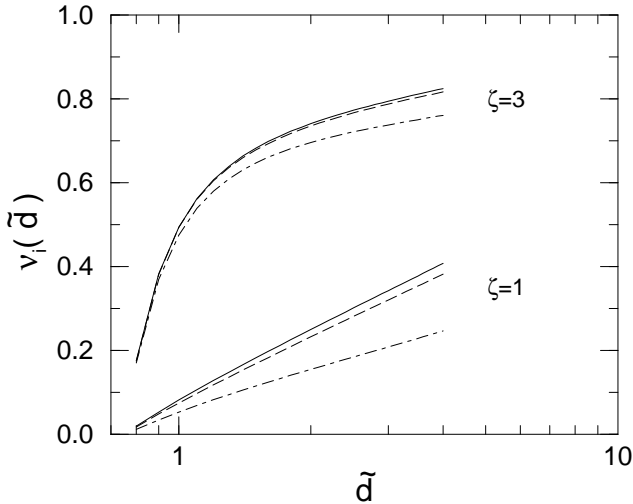}
}
\caption{Dependence of  $\nu_i(\tilde{d})$ on the number of arms $f$ and 
boundary conditions (size $\tilde{A}$ of the cubic cell). Full lines: $N=130$,
$f=27$ and $\tilde{A}=320$; 
 dashed lines: $N=130$, $f=27$ and $\tilde{A}=640$ ;
dash-dotted lines: $N=130$, $f=6$ and $\tilde{A}=640$.}
\label{fig:5}       % Give a unique label
\end{figure}

The Fig.~15 represents the density $n(\tilde{r},\tilde{d})$ 
of counterions at distance
 $d=\tilde{d}a$ from the closest branch and at a distance 
$\tilde{r}a$  from the center of the micelle for three values
of $\tilde{r}$: close to the core $(\tilde{r}=20)$, 
in the middle 
 $(\tilde{r}=40)$ and close to the edge of the corona   
$(\tilde{r}=70)$. We chose
$\zeta=3$, well above the Manning threshold, 
$N=130$, $f=27$ and $\tilde{A}=320$.

It is interesting to compare these results to those expected from
a ``rod-like'' picture of the arms and a Poisson-Boltzmann approach
for the distribution of counterions.

Using the ``FIDISOL/CADSOL'' package developed at
Rechenzentrum, Universit\" at Karlsruhe \cite{fidisol} for the numerical
solution of non-linear partial differential equations, we have
solved the Poisson-Boltzmann equation for a rod in a cone of solid
angle $(4\pi/f)$. The radius of the rod is chosen to $0.75a$, which
corresponds to the sum of the hard-core radius of monomer and counter ion. 
The height of the cone is equal to the radius of a sphere of volume $A^3$.
The distribution of charges in the rod corresponds
to the monomer profile obtained in Fig.~3c through Monte-Carlo
simulation. (In practice, we  fit this distribution to an
approximation by a rational function).
Using a two-dimensional $80\times 120$ grid, we  obtained an 
accuracy better than $10^{-4}$ on almost every point.
The calculation took
only one minute of monoprocessor time. There is a nice agreement except
in the vicinity of the corona radius.

Fig.~16 compares the overall density profile $\rho_i (\tilde{r})$ obtained
through the Poisson-Boltzmann approach to the Monte-Carlo results
of Fig.~3c.

An interesting quantitative way to characterize 
the Manning threshold is to plot
in log-linear coordinates the number $\nu_i(\tilde{d})$ of counterions at
distance less than $\tilde{d}$ from the closest monomer (i.e 
the integral over $\tilde{d'}$ of  
$n (\tilde{r},\tilde{d'})$ between $3/2$ and $\tilde{d}$) \cite{Luc2}.
Within the Poisson-Boltzmann approximation, 
$\nu_i(\tilde{d})$ is expected
to follow a logarithmic law at the Manning threshold \cite{PB}.
As  shown in Fig.~17, there is a change of curvature
of the lines between $\zeta=0.5$ and $\zeta=1.5$. A straight line
(logarithmic behavior) is obtained at $\zeta=1$.

The Fig.~18 shows that,
in the Manning condensed regime, $\nu_i(\tilde{d})$ 
is practically independent of
the boundary conditions: the results
obtained after doubling the size of the cubic cell ($\tilde{A}=640$) are 
practically identical to those obtained with $\tilde{A}=320$. 
$\nu_i(\tilde{d})$ depends on the number of arms
 but the threshold value $\zeta_M\approx 1$ 
corresponding to Manning condensation is unchanged.

\begin{figure}
\resizebox{0.75\textwidth}{!}{%
  \includegraphics{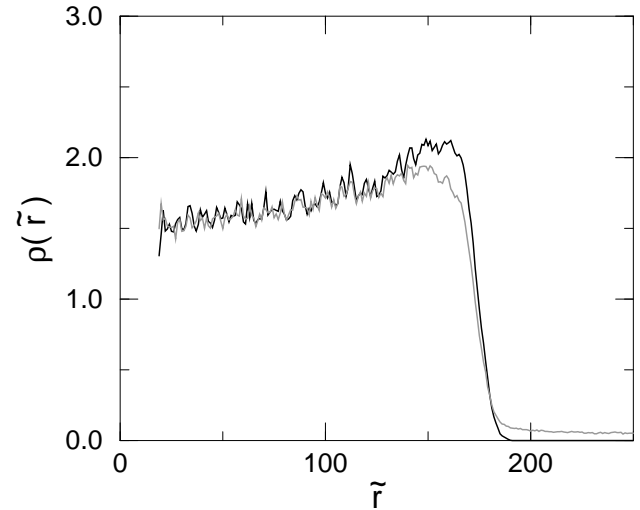}
}
\caption{Density profiles of monomers (black line) and counterions
(grey line) for a micelle with 54 branches of 251 monomers at 
$\zeta = 2.8$ with $R_{core}=18a$ and $\tilde{A}=560$.} 
\end{figure}

\begin{figure}
\resizebox{0.75\textwidth}{!}{%
  \includegraphics{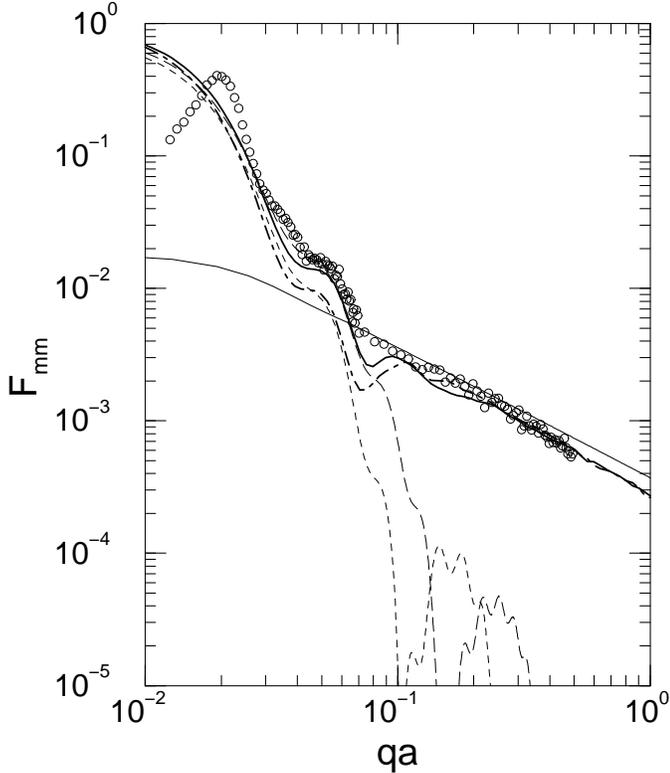}
}
\caption{Monomer-monomer form factor
for a micelle with 54 branches of 251 monomers at $\zeta = 2.8$.
The experimental results (circles) are compared to two Monte-Carlo
calculation with different core size (all other parameters being the same)
(i) heavy full line: $R_{core}=12a$; (b) heavy dash-dotted line 
 $R_{core}=18a$. The thin lines indicate analytical asymptotic behaviors.
Thin full line: the form factors of perfect rods of length $R_c-R_{core}$;
Thin dashed and dotted lines the form factors for spherical distributions
of monomers in $1/r^2$ between $R_{core}$ and $R_c$ for $R_{core}=12a$ and
$R_{core}=18a$ respectively.}
\end{figure}

\section{Simulation of an experimental system}

\subsection{Description of the experimental system}

We used asymmetric diblock copolymers made of fully deuterated sodium
poly-(styrene-sulfonate) $({\rm NaPSS_d})$ and of
(polyethylene-propylene) (PEP) \cite{Muller1}. These materials 
were synthetized by anionic polymerization.
The degrees of polymerization were measured and found to be 251 for 
$({\rm NaPSS_d})$ moiety and 52 for PEP moiety. The aggregation number,
i.e the number of branches per micelle, was found to be 54 and the core radius
is  $R_{core}\approx 4.5~nm\approx 18a$.
The monomer size for   $({\rm NaPSS_d})$ moieties is $a=0.25 nm$. Taking
$l_B=0.71 nm$ as the Bjerrum length of pure water at room temperature,
the dimensionless parameter $\zeta$ is evaluated to $\zeta=l_B/a=2.8$.
 The concentration
is $0.25 c^*$ where $c^*$ represents the overlap concentration.
Our cubic cell was chosen such that its volume represents the mean volume
occupied by one micelle in the solution i.e. $\tilde{A}=560$.

\subsection{Monomer and counterion profiles}

The density profiles of monomers and counterions are plotted in 
Fig.~19. The corona radius is $R_c=174a$ leading to 
$(R_c-R_{core})/Na \approx 0.62$ this value has to be compared to
that obtained in section 3 with $N=130$ and $f=54$ at the same $\zeta$:
$(R_c-R_{core})/Na \approx 0.57$. We  agree with the Molecular
Dynamics results of JLL \cite{Likos1,Likos2} on smaller branch lengths showing
that the ratio $R_c/N$ is not strictly constant but slightly decreases
when $N$ increases. The monomer and ion profiles are quite similar
to those observed in section 3 for $N=130$ and $f=27$ (compare with
Fig.~3c).

\subsection{Form factor, comparison to experimental results}

The monomer-monomer form factor obtained through the Monte-Carlo
simulation, with the parameters described in   subsection 4.1
is shown in Fig.~20 (dashed line). It is  close
to the monomer-monomer form factor obtained though neutron
scattering \cite{Muller1} except in the range $q\approx \pi/R_{core}$ 
where the 
size of the core has a large influence on the spectrum.
Another Monte-Carlo calculation with a slightly smaller
core $R_{core}=12a$, (full line) was performed. It is in a better 
agreement with the 
neutron scattering results. 

It is interesting to compare the simulated and experimental results 
to straightforward
analytical approximations at low and high $q$. 

\begin{enumerate}
\item[(i)]
At low $q$, 
($ q < \pi/R_{core}$), the form factor can be reasonably approximated
by that of a $1/r^2$ spherical distribution of monomers from $r=R_{core}$
to $r=R_c$ and zero otherwise:
\begin{equation}
F^{sph}(q)= \left[ {Si(qR_c)-Si(qR_{core})\over qR_c}\right]^2 
\end{equation}
where $Si(z)$ represents the SineIntegral function \cite{Stegun}
The thin dashed and dotted lines represent this function for 
$R_{core}=12a$ and
$R_{core}=18a$ respectively. Up to $qa\approx 0.1$ they closely follow
the corresponding Monte-Carlo simulations (heavy full and dashed lines).
At $q\approx \pi/R_{core}$ large oscillations appear in the function 
$-Si(qR_{core})$, the first of them being responsible for a dip observed in
the simulation. Close to this dip, the form factor is very sensitive to
the size of the core. We find that our simulation with $R_{core}=12$ is
closer to the experimental results than that with $R_{core}=18$. 
The interface between the hydrophobic core and the hydrophilic branches
might happen to be not so smooth and as  well defined as assumed in our model.
For example,  some hydrophilic monomers may be  trapped inside the outer
part of the core, leading to a lower effective core radius.

\item[(ii)]
At high $q$, the branches can be approximated by rods of length
 $L=R_c-R_{core}$
with negligible thickness. The corresponding form factor is:
\begin{equation}
F^{rod}(q)={1\over f}
\left\{{2\over qL}Si(qL)-\left[{\sin(qL/2)\over qL/2}\right]^2\right\}
\end{equation}
The thin full line represents this function with $L=R_c-R_{core}=156a$
corresponding to our Monte-Carlo estimate. At large $q$,
$F^{rod}(q)\approx [1/fq(Rc-R_{core})]$ and the position of this 
line of slope (-1) in logarithmic coordinates gives a nice estimate of
the corona radius.
In conclusion the corona radius obtained through our Monte-Carlo simulation
is in agreement with the experimental value.
\end{enumerate}
The peak observed in the experimental data at $qa\approx 0.02$ arises
from long-range correlations between micelles, mostly due to the
remaining uncondensed charges. It is a collective effect
which, of course, is absent in our model. The concentration is 
$c=5\times 10^{-3}g/cm^3$. The corresponding $q$ wave vector characterizing
the inter-micelle correlations is $q\approx 2\pi [c{\cal{N}}/fM]^{1/3}$
where $M=5.7\times 10^4$ is the molar mass of one diblock copolymer 
and $\cal{N}$ the Avogadro
number. We obtain $qa\approx 0.016$ in  reasonable agreement with
the position of the peak.

\section{Conclusion}

Monte-Carlo simulations appear to be a useful tool for 
understanding the complex
structure of highly charged copolymer urchin-like micelles. As the 
Coulomb coupling is increased, three different regimes can be distinguished.
i) At low coupling, the freely-jointed polyelectrolyte branches become
stretched due to the monomer-monomer repulsion and the micelle swells. The
monovalent counterions are not sensitive to the low micellar charge and are 
uniformly distributed inside the cell. ii) At intermediate coupling, the arms
are almost fully stretched and most of the counterions are trapped inside or 
in close vicinity of the micelle. This spherical condensation is due to the
presence of many, although weakly charged, branches in the same micelle.
iii) At large coupling. a classical ionic condensation appears 
locally around
each highly charged branch which adds to the previous one. The effective
post-condensation monomer charge decreases, the monomer-monomer repulsion
 weakens and the micelle deswells. These Monte-Carlo results agree
with Poisson-Boltzmann analyses in different geometries and nicely
reproduce small-angle neutron scattering data. 
The influence of intrinsic chain rigidity, 
and the role of ion-ion correlation in presence of multivalent counterions
will  be investigated in future papers.



\begin{thebibliography}{}
\bibitem{Footnote1}
JLL consider $b=R_c/(N\alpha)$ rather than $b=a/\alpha$. The relevant
question is: how to extend the Manning concepts from a perfect
cylindrical rod to a wiggling chain. We consider that the cylindrical
picture remains valid at small length scale (of the order of 
ten  monomers) and that our local definition $b=a/\alpha$ is more
relevant that the global one  $b=R_c/(N\alpha)$, which incorporates the
large scale wiggles of the chain shown in Fig.~13 (see  footnote [23]).
Nevertheless taking  $b=R_c/(N\alpha)$ instead of $b=a/\alpha$ merely
shifts in Fig.~10 the data of JLL as well as ours from a factor $Na/R_c$ 
and does not affect our general discussion.
\bibitem{Footnote2}
This blob picture in the ``Manning
condensed'' regime, with effective charge renormalization, is identical to
that proposed by Schiessel and Pincus [24] for dilute solutions
of polyelectrolytes, except that we consider a renormalizing factor:
$
\alpha_{eff}=1/\zeta=a/l_B
$
while their equation (2) corresponds, with our notations, to
$
\alpha^{SP}_{eff}=(R_c/N)/l_B
$
and leads to the exponent -1 instead of (-1/3).
We are left with the same question as the one risen in  footnote [21]
about how to define the  ``Manning parameter'' for an imperfect rod.
In Fig.~10, our Monte Carlo results for $R_c$ show
a (-1/3) exponent. This indicates that our ``local'' picture is the most
relevant.
\end{thebibliography}
\end{document}